\documentclass[superscriptaddress,amsmath,amssymb,aps,pra, twocolumn]{revtex4-2}

\usepackage{graphicx}% Include figure files
\usepackage{dcolumn}% Align table columns on decimal point
\usepackage{bm}% bold math
\usepackage[colorlinks]{hyperref}% add hypertext capabilities

\newcommand{\dlangle}{\langle \langle}
\newcommand{\drangle}{\rangle \rangle}

\begin{document}

\title{Dynamical nonlocality in quantum time via modular operators}

\author{I. L. Paiva}
\affiliation{Faculty of Engineering and the Institute of Nanotechnology and Advanced Materials, Bar-Ilan University, Ramat Gan 5290002, Israel}

\author{M. Nowakowski}
\affiliation{Faculty of Applied Physics and Mathematics, Gdansk University of Technology, 80-952 Gdansk, Poland}
\affiliation{National Quantum Information Center of Gdansk, Andersa 27, 81-824 Sopot, Poland}

\author{E. Cohen}
\affiliation{Faculty of Engineering and the Institute of Nanotechnology and Advanced Materials, Bar-Ilan University, Ramat Gan 5290002, Israel}

\begin{abstract}
We formalize the concept of the modular energy operator within the Page and Wootters timeless framework. As a result, this operator is elevated to the same status as the more studied modular operators of position and momentum. In analogy with dynamical nonlocality in space associated with the modular momentum, we introduce and analyze the nonlocality in time associated with the modular energy operator. Some applications of our formalization are provided through illustrative examples.
\end{abstract}

\maketitle

\section{Introduction}

Nonlocality is a remarkable concept that has been attracting an ever-increasing interest from the community since the early days of quantum theory \cite{einstein1935can, schrodinger1935discussion} till now. Whether it is through Bell's nonlocality \cite{bell1964einstein, brunner2014bell}, quantum steering \cite{wiseman2007steering, uola2020quantum}, quantum entanglement in general \cite{horodecki2009quantum}, or, even more broadly, quantum discord \cite{zurek2000einselection, ollivier2001quantum, henderson2001classical, modi2012classical}, the topic has been central in the studies of quantum foundations, and with a good reason: since multiple experiments verified the quantum violation of Bell's inequalities \cite{freedman1972experimental, clauser1978bell, aspect1981experimental, aspect1982experimental, aspect1982experimental2, genovese2005research, shalm2015strong, giustina2015significant}, it is believed that quantum mechanics is fundamentally different from classical mechanics. These studies have led to theoretical and technological breakthroughs \cite{ekert1991quantum, bennett1992communication, bennett1993teleporting, vaidman1994teleportation, braunstein1998teleportation, jozsa2000quantum, lesovik2019arrow, micadei2019reversing, arute2019quantum}. Moreover, it is even possible to discuss \textit{entanglement in time} \cite{taylor2004entanglement, cotler2015bell, nowakowski2017quantum, cotler2017experimental, nowakowski2018}.

The above type of nonlocality is associated with the preparation (or preparation and measurement) of systems. It can, then, be referred to as \textit{kinematic nonlocality}. A different type of nonlocality, which was introduced using the concept of modular variables \cite{aharonov1969modular}, is associated with the equations of motion obeyed by quantum systems and, hence, called \textit{dynamical nonlocality}. Although very promising, as already shown in first applications to quantum information with continuous systems \cite{gottesman2001encoding, gneiting2011detecting, vernaz2014continuous, ketterer2016quantum}, these variables have not fully received the corresponding attention from a significant part of the community yet \cite{popescu2010dynamical}.

The most common types of modular variables considered in the literature are the modular position and the modular momentum \cite{gottesman2001encoding, massar2001greenberger, aharonov2005quantum, tollaksen2009deterministic, tollaksen2010quantum, popescu2010dynamical, gneiting2011detecting, carvalho2012experimental, vernaz2014continuous, lobo2014weak, barros2015detecting, ketterer2016quantum, aharonov2017finally, fluhmann2018sequential}. In fact, letting $\ell$ and $p_0$ be parameters with dimensions of length and momentum, respectively, the modular operators
\begin{equation}
    e^{iXp_0/\hbar}
    \label{eq:exp-mod-x}
\end{equation}
and
\begin{equation}
    e^{iP\ell/\hbar},
    \label{eq:exp-mod-p}
\end{equation}
associated with the modular position and momentum, are studied in various scenarios, specially in interference effects. However, other types of modular variables are also considered. In the next section, further background regarding these operators is provided.

For this work, modular energy, which is related to dynamical nonlocality in time, is of particular interest. Formally, it does not have the same status as the other modular variables. While, say, a modular position is ``naturally'' associated with a modular momentum, modular energy does not find an associated modular time because time is a parameter, and not an observable, in quantum mechanics. Still, it is known from relativity theory that momentum and energy are parts of a single property (the stress-energy tensor) of physical objects in spacetime. This line of reasoning, then, leads to the concept of modular energy, given by
\begin{equation}
    e^{iH_{S}\tau/\hbar},
    \label{mod-energy}
\end{equation}
where $\tau$ is a parameter with units of time and $H_S$ is the Hamiltonian of the system of interest, was also introduced. It is called a modular operator since the time parameter $\tau$ defines different modular energies and is associated with the energy mod $2\pi\hbar/\tau$. Although the operator in Eq. \eqref{mod-energy} is familiar from the study of quantum dynamics, as a modular variable, it is assumed to be a property of the system, which makes it more delicate conceptually compared to the modular variables of position and momentum, for instance. This is the case, in part, since time, the canonically conjugate variable of the Hamiltonian $H$, is not an observable in quantum mechanics, as already mentioned. Nevertheless, the time evolution of the modular momentum, which is displayed in Eq. \eqref{nonloc-space} and involves a dynamical notion of nonlocality in space, seems to suggest the existence of \textit{nonlocality in time} for the evolution of the modular energy, i.e., the time derivative of the modular energy might, in general, depend on temporally remote events. It is this notion of temporal nonlocality that we wish to investigate in the current work.

This is a subject that may play a relevant role in the study of phenomena with some periodicity in time, like time crystals \cite{wilczek2012quantum, wilczek2013superfluidity, bruno2013comment, bruno2013impossibility, watanabe2015absence, khemani2016phase, else2016floquet, von2016absolute, yao2017discrete, zhang2017observation} and many others, as is further discussed later in this work. Moreover, it may lead to new discoveries, including novel insights into conservation laws in quantum mechanics \cite{aharonov2021conservation}.

In our analysis, envisioning a better formalization of the nonlocality in time, we employ a previously suggested framework of \textit{timeless quantum mechanics}. Particularly relevant here is the approach introduced by Page and Wootters \cite{page1983evolution}, which has recently been subjected to much scrutiny \cite{giovannetti2015quantum, giacomini2019quantum, smith2019quantizing, diaz2019history, diaz2019history2, castro2020quantum, smith2020quantum}. According to this approach, a (quantum) clock system is used as a reference for the evolution of the system of interest. Together, the clock and the system of interest are often assumed to be a closed system. It can be shown that the evolution of the main system with respect to the time given by the clock is typically unitary -- even in a scenario with multiple clocks \cite{castro2020quantum}.

Hereon, we shall study the modular energy operator using the framework of timeless quantum mechanics. First, in the next section, we review some basic properties of modular variables. Then, in Section \ref{sec:mod-energy}, we introduce the Page and Wootters timeless framework for quantum mechanics as well as our formalization of modular energy within it. Following that, in Section \ref{sec:operational}, we discuss the operational meaning of the theoretical objects introduced in our study. In Section \ref{sec:applications}, we present and analyse two applications of it. Finally, in Section \ref{sec:discussion}, we present our last remarks.

\section{Modular variables and some of their properties}
\label{appendix}

As mentioned in the Introduction, two of the most common modular variables considered in the literature are the modular position and modular momentum, given, respectively, by
\begin{equation}
    X_\text{mod} \equiv X \mod 2\pi\hbar I/p_0
    \label{eq:def-mod-x}
\end{equation}
and
\begin{equation}
    P_\text{mod} \equiv P \mod 2\pi\hbar I/\ell,
    \label{eq:def-mod-p}
\end{equation}
where $I$ is the identity operator, $X$ and $P$ are the usual position and momentum operators, and $\ell$ and $p_0$ are parameters with dimensions of length and momentum, respectively. The adjective \textit{modular} in modular variables comes from the periodicity inherent to them. As it will be better explained in this Introduction, although the parameters $\ell$ and $p_0$ can be, in principle, arbitrary, there is typically a ``natural'' choice for them.

These concepts were introduced as a tool in the study of interference in the Heisenberg picture \cite{aharonov2005quantum, tollaksen2009deterministic, aharonov2017finally}. More specifically, some of these variables consist of (nonlocal) properties of a particle whose expectation values may be functions of the relative phase between two coherent wavepackets -- even while they are spatially separated. In particular, if a system is in the state
\begin{equation}
    |\psi(t)\rangle = \frac{1}{\sqrt{2}} \left(|\xi_1(t)\rangle + e^{i\varphi}|\xi_2(t)\rangle\right),
    \label{eq:state-with-phase}
\end{equation}
where $\xi_1$ and $\xi_2$ are two wavepackets, it is possible to show that the expectation value of any power of $X$ and $P$ or products thereof (i.e., any polynomial operator of the form $f(X, P)=\sum_{mn} a_{mn} X^{m}P^{n}$, where $a_{mn}\in\mathbb{C}$ for every $m,n$) does not depend on $\varphi$ if $\xi_1$ and $\xi_2$ are orthogonal to each other \cite{aharonov2005quantum}. However, the modular operators in Eqs. \eqref{eq:exp-mod-x} and \eqref{eq:exp-mod-p}, which are equivalent to $e^{iX_\text{mod}p_0/\hbar}$ and $e^{iP_\text{mod}\ell/\hbar}$, respectively, may include $\varphi$ in their expectation values. It should be noticed that, while the operators defined in Eqs. \eqref{eq:exp-mod-x} and \eqref{eq:exp-mod-p} are non-Hermitian, observables can be easily defined from them.

To illustrate that, let $\langle x|\xi_1\rangle \equiv \xi(x)$ and $\langle x|\xi_2\rangle \equiv \xi(x-\ell)$, for some $\xi(x)$ with small enough support around one of the slits. Then, it can be verified by direct computation that
\begin{equation}
    \langle e^{iP\ell/\hbar}\rangle = \frac{1}{2} e^{i\varphi}.
    \label{eq:expval-phase}
\end{equation}
This example can be understood, for instance, as the analysis of the double-slit experiment, where $\ell$ is the separation between the slits.

\begin{figure}
    \centering
    \includegraphics[width=\columnwidth]{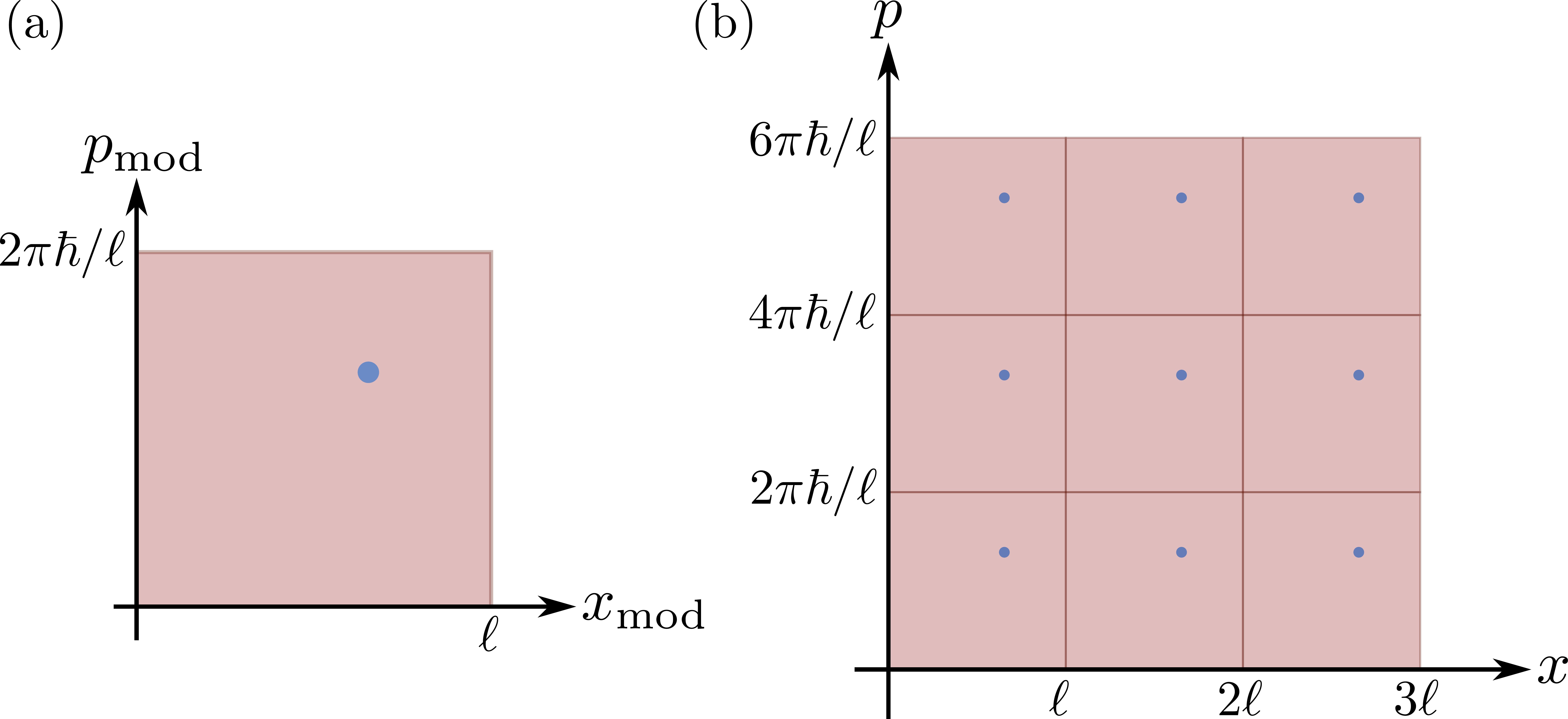}
    \caption{\textbf{Representation of the modular and the usual phase spaces.} (a) Since modular position and modular momentum commute, the system's state can be represented by a single point in the \textit{modular phase space}. This seems to violate Heisenberg's uncertainty principle. (b) However, the two ideas can be reconciled by observing that the usual phase space is divided into periodic cells. Although the system must be represented by a single point on each cell, the modular variables do not provide any information about the cell to which the state of the system belongs.}
    \label{fig:phase-space}
\end{figure}

Like any property, modular variables are bound to a conservation law \cite{aharonov2005quantum}. However, differently from the standard position and momentum, it can be checked that, for $p_0=2\pi\hbar/\ell$, the modular position and momentum defined in Eqs. \eqref{eq:def-mod-x} and \eqref{eq:def-mod-p} commute with each other, a property that was recently experimentally investigated \cite{fluhmann2018sequential}. While that seems to violate the uncertainty principle, this is not the case. The uncertainty principle is manifested in a different manner \cite{aharonov2005quantum}. Specifically, $P_\text{mod}$ and $X_\text{mod}$ divide the phase space into periodic cells. Their commutation means that, for a given state, a point can be assigned to each cell. However, $P$ and $X$ remain unknown since their modular counterparts do not provide information on which cell is associated with the state. This characteristic is illustrated in Fig. \ref{fig:phase-space}. Also, it is possible to introduce a \textit{complete uncertainty relation} for modular variables \cite{aharonov2005quantum, aharonov2017finally}, which states that, for any (dimensionless) modular variable $\Phi_\text{mod}\equiv\Phi\mod 2\pi I$, the expected value $\langle e^{in\Phi}\rangle$ vanishes for every integer $n$ if and only if the modular variable $\Phi_\text{mod}$ is completely uncertain. In fact, this follows from the Fourier series expansion of the probability $P_{\Phi_\text{mod}}$ of $\Phi_\text{mod}$ assuming an arbitrary value $\varphi \in [0,2\pi)$, i.e.,
\begin{equation}
    P_{\Phi_\text{mod}}(\varphi) = \sum_{n\in\mathbb{Z}} c_n e^{in\varphi},
\end{equation}
where
\begin{equation}
    c_n = \frac{1}{2\pi} \int_0^{2\pi} P_{\Phi_\text{mod}}(\varphi) e^{in\varphi} d\varphi = \frac{1}{2\pi} \langle e^{in\Phi}\rangle.
\end{equation}
Then, $\Phi_\text{mod}$ is completely uncertain, i.e., $P_{\Phi_\text{mod}}(\varphi)$ is a uniform distribution for every $\varphi \in [0,2\pi)$ if and only if $\langle e^{in\Phi}\rangle$ vanishes for every non-zero integer $n$.

Fundamentally, these variables evolve in a nonlocal manner. To see that, let $H=P^2/2m+V(X)$ be the Hamiltonian of the system of interest, where $m$ is the mass of the system and $V$ is the potential. Then, it follows from Heisenberg's equation of motion that
\begin{equation}
    \begin{aligned}
    \frac{d}{dt}e^{iP\ell/\hbar} & = -\frac{i}{\hbar} \left[e^{iP\ell/\hbar}, H\right] \\
    & = -\frac{i}{\hbar} \left[V(X+\ell I)-V(X)\right] e^{iP\ell/\hbar},
    \end{aligned}
    \label{nonloc-space}
\end{equation}
which depends on spatial locations separated by a length $\ell$. The fundamental observation is that information about the distance $\ell$ is coded in the modular momentum via $p_0=2\pi\hbar/\ell$ for $p_{mod}=p \mod p_{0}$. This is in contrast with the classical equation of motion, which leads to
\begin{equation}
    \begin{aligned}
    \frac{d}{dt}e^{2\pi i p/p_0} &= \left\{e^{2\pi i p/p_0}, H\right\} \\
    &= -i \frac{2\pi}{p_0} \frac{dV}{dx} e^{2\pi i p/p_0},
    \end{aligned}
    \label{clas-evol-mod-momentum}
\end{equation}
i.e., the time evolution of the modular momentum depends on a (local) derivative with respect to the spatial variable.

To give a concrete example, we can refer to the double-slit experiment again. As discussed in Refs. \cite{aharonov2005quantum, aharonov2017finally}, Eq. \eqref{nonloc-space} with $\ell$ given by the distance between the slits allows an interpretation where the particle goes through one slit but has its modular momentum affected by a potential at both slits. Moreover, if a measurement is performed at the slits in order to determine the path taken by the particle, the information about its modular momentum is destroyed in this picture. In fact, for the choices of $|\xi_1\rangle$ and $|\xi_2\rangle$ in Eq. \eqref{eq:state-with-phase} that lead to the expected value associated with $|\psi\rangle$ in Eq. \eqref{eq:expval-phase}, it can be noted that, in spite of the latter equation, $\langle e^{iP\ell/\hbar}\rangle$ vanishes for the states $|\xi_1\rangle$ and $|\xi_2\rangle$. Then, when their superposition is destroyed by a measurement, the modular momentum of the particle becomes completely uncertain. This, in turn, preserves relativistic causality.

At this point, we are in a good position to justify the choice of the parameter $\ell$ as the distance between the slits, something not explained in the literature prior to this work. If $\ell$ is much bigger than the separation between the slits, then $\langle e^{iP\ell/\hbar}\rangle$ also vanishes for the state $|\psi\rangle$, i.e., this modular variable remains uncertain throughout the experiment and, then, is not relevant in the analysis of it. When $\ell$ approaches the length of separation of the slits, $\langle e^{iP\ell/\hbar}\rangle$ becomes non-zero. Although they are, then, relevant, they do not provide as much information as the choice of $\ell$ coinciding with the separation between the slits. This is the sense in which we said that there is typically a natural choice for the parameters associated with a modular variable.

\section{Nonlocality in time within the timeless framework}
\label{sec:mod-energy}

The timeless framework \cite{page1983evolution} considers a clock system, whose state is given by a vector in a Hilbert space $\mathcal{H}_A$, and the system whose evolution is studied, represented by a state in a Hilbert space $\mathcal{H}_R$, where $R$ stands for the ``rest.'' The joint system $|\Psi\drangle \in \mathcal{H}_A\otimes\mathcal{H}_R$ is assumed to be closed and, hence, it is subject to the Wheeler-DeWitt equation,
\begin{equation}
    H_T|\Psi\drangle = 0,
    \label{constrain-eq}
\end{equation}
where $H_T$ is the total Hamiltonian acting on systems $A$ and $R$. The double ket notation is used to denote the full isolated system.

Now, let $T_A$ be the time operator associated with clock $A$ and $H_A$ its free Hamiltonian, with $[T_A,H_A]=i\hbar$, which implies that $H_A=-i\hbar \partial/\partial t_A$. Observe that $T_A$ and $H_A$ are the analogous in $\mathcal{H}_A$ to the usual $X$ and $P$ in standard quantum mechanics. Also, let $H_R$ be the free Hamiltonian of the system of interest, and let $H_{int}(T_A)$ represent the interaction between $A$ and $R$, which is analogous to the time-dependent terms of the evolution of $R$ in standard formulations of quantum mechanics. Then,
\begin{equation}
    H_T = H_A + H_R + H_{int}(T_A).
    \label{eq:ht}
\end{equation}
Replacing it in Eq. \eqref{constrain-eq} and applying a scalar product by an eigenstate $|t_A\rangle$ of $T_A$ on the left, i.e., $|\psi(t_A)\rangle=\langle t_A|\Psi \drangle$, which results in a reduced state of $R$ conditioned on the state $|t_A\rangle$ of the clock $A$, it holds that
\begin{equation}
    i\hbar \frac{\partial}{\partial t_A} |\psi(t_A)\rangle = \left[H_R + H_{int}(t_A)\right]|\psi(t_A)\rangle,
    \label{eq:schrod}
\end{equation}
which is the Schr\"odinger equation for the system $R$ with time measured by the external clock $A$. As a result, $|\Psi\drangle$ can be written as
\begin{equation}
    |\Psi\drangle = \int dt_A \ |t_A\rangle \otimes |\psi(t_A)\rangle,
    \label{eq:history}
\end{equation}
where $|\psi(t_A)\rangle$ is the usual (normalized at each instant of time) state vector considered in quantum mechanics. Because $|\Psi\drangle$ contains information about $|\psi(t_A)\rangle$ at every $t_A$, it is referred to as the \textit{history state}.

Observe that, since the system composed of $A$ and $R$ is assumed to be isolated, clock $A$ mediates the interaction of any other system in the universe with $R$, as already discussed in Ref. \cite{smith2019quantizing}. In particular, any change in the energy distribution of $R$ generated by the Hamiltonian in Eq. \eqref{ta-hamiltonian} corresponds to a change in the energy distribution of clock $A$.

\begin{figure}
    \centering
    \includegraphics[width=\columnwidth]{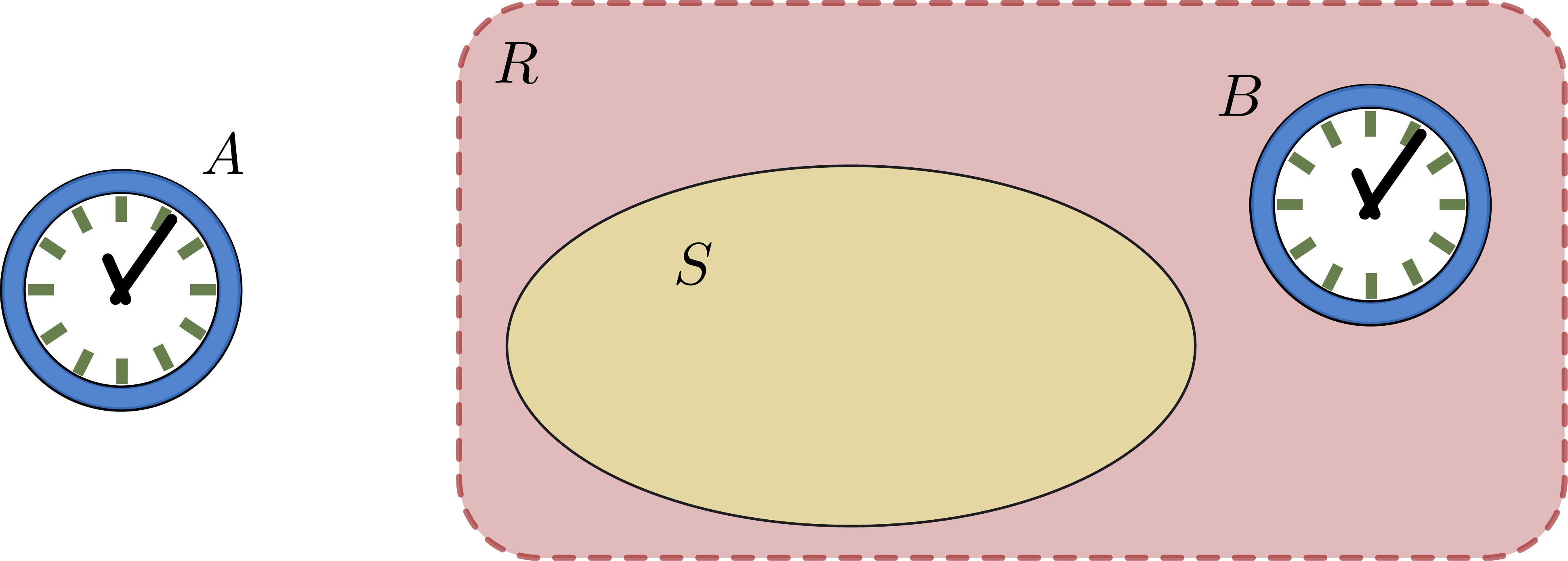}
    \caption{\textbf{Scheme of the configuration considered in this work.} An external clock represents the interaction of the system of interest $R$ with its exterior. System $R$ itself is composed of an internal clock $B$ and a main system $S$.}
    \label{fig:scheme}
\end{figure}

Here, we consider the case studied in Ref. \cite{castro2020quantum}, where system $R$ was assumed to be composed of the main system of interest $S$ and a clock $B$, i.e., an internal clock to $R$, as represented in Fig. \ref{fig:scheme}. We, then, write
\begin{equation}
    H_R = H_B + H_S + H_{BS}(T_B),
\end{equation}
where $H_{BS}(T_B)$, assumed to be such that $[H_B, H_{BS}(t)]=0$ for a parameter $t$, generates the inner unitary transformation of system $S$ controlled by the time in clock $B$, i.e., changes on system $R$ when it is completely isolated. In this case, Eq. \eqref{eq:ht} and, as a consequence, Eq. \eqref{eq:schrod} hold.

For simplicity, the term $H_{BS}(T_B)$ is assumed to be null in this work. This implies that the effective Hamiltonian of system $R$ from the perspective of clock $A$ is
\begin{equation}
    H_{eff}^A = H_B + H_S + H_{int}(t_A).
    \label{ta-hamiltonian}
\end{equation}
We also restrict to cases where $[H_B,H_{int}(T_A)]=0$, i.e., $H_{int}$ is not a function of $T_B$.

Our goal is to obtain the dynamics of the systems from the perspective of clock $B$, although a similar conclusion can be drawn based on an analysis from the perspective of clock $A$. Then, writing $\langle t_B|\Psi\drangle = |\varphi(t_B)\rangle \in \mathcal{H}_A \otimes \mathcal{H}_S$, it holds that
\begin{equation}
    i\hbar \frac{\partial}{\partial t_B} |\varphi(t_B)\rangle = [H_A + H_S + H_{int}(T_A)] |\varphi(t_B)\rangle,
\end{equation}
i.e., from $B$'s perspective, the effective Hamiltonian is
\begin{equation}
    H_{eff}^B = H_A + H_S + H_{int}(T_A)
\end{equation}

From the Heisenberg equation of motion, we obtain the $t_B$-evolution, i.e., the evolution from the perspective of the clock $B$ of the modular energy $e^{-iH_A\tau/\hbar}$ of clock $A$, where $\tau$ is a parameter with units of time. Explicitly,
\begin{equation}
    \begin{aligned}
    \frac{d}{dt_B}e^{iH_A\tau/\hbar} &= -\frac{i}{\hbar} \left[e^{iH_A\tau/\hbar}, H_{eff}^B\right] \\
    &= -\frac{i}{\hbar} \left[e^{iH_A\tau/\hbar} H_{int}(T_A) - H_{int}(T_A) e^{iH_A\tau/\hbar}\right] \\
    &= -\frac{i}{\hbar} \left[H_{int}(T_A + \tau I) - H_{int}(T_A)\right] e^{iH_A\tau/\hbar},
    \end{aligned}
    \label{mod-energy-nonlocality}
\end{equation}
which is to be compared with Eq. \eqref{nonloc-space}. Observe that no particular set of assumptions was made in this derivation. In fact, the result follows from the fact that $e^{iH_A\tau/\hbar}$ generates translations in the time variable of clock $A$ and $H_{eff}^B$ is, in general, a function of $T_A$. Also, although Eq. \eqref{mod-energy-nonlocality} seems to allow backwards-in-time signaling, we recall that this is not the case because of the already discussed complete uncertainty principle. Moreover, in a study from the perspective of clock $A$, the evolution of the modular energy of clock $B$ would, in general, depend on the effective Hamiltonian at two different points in time. For that, one should take into consideration $H_{BS}(T_B)$, which is only assumed to vanish in this work for simplicity.

The result of Eq. \eqref{mod-energy-nonlocality} becomes particularly interesting when two other results are pointed out. The first is similar to the classical evolution of the modular momentum represented in Eq. \eqref{clas-evol-mod-momentum}. In fact, since $\partial e^{2\pi iE_A/E_0}/\partial t_A$ vanishes, the classical dynamical equation for the modular energy leads to
\begin{equation}
    \begin{aligned}
    \frac{d}{dt_B}e^{2\pi iE_A/E_0} &= \left\{e^{2\pi iE_A/E_0}, H_{eff}^B\right\} \\
    &= - \frac{\partial e^{2\pi iE_A/E_0}}{\partial E_A} \frac{\partial H_{eff}^B}{\partial t_A} \\
    &= -i \frac{2\pi}{E_0} \frac{dH_{int}}{dt_A}e^{2\pi iE_A/E_0},
    \end{aligned}
\end{equation}
which, again, depends only on (local) derivatives.

The second result concerns the fact that Eq. \eqref{mod-energy-nonlocality}, in principle, contains two different notions of time, from clocks $A$ and $B$. This equation displays dynamical nonlocality in time $B$ in a similar manner that Eq. \eqref{nonloc-space} displays dynamical nonlocality in $X$. However, this does not necessarily correspond to the usual idea of dynamical nonlocality in time discussed in the literature. To recover this, it suffices to show an equivalence of the notion of time provided by clock $A$, associated with the nonlocality on the right-hand side of the equation, and clock $B$, which provides the time used as reference for the analysis of the evolution. This is indeed the case, as follows from the Heisenberg equation of motion
\begin{equation}
    \frac{d}{dt_B} T_A = -\frac{i}{\hbar} [T_A,H_{eff}^B] = I.
    \label{eq:time-rel}
\end{equation}
This means that, in our model, the ``flow of time,'' when described in terms of the rate of change of the mean value shown by a clock, is the same in both clocks. Nevertheless, as evidenced by Eq. \eqref{mod-energy-nonlocality}, with respect to system $R$'s proper time, the dynamics of the modular energy of the external system $A$ depends on a future instant of time.

Even if one introduces a different notion for the flow of time that includes the variance of $T_A$, our conclusion still holds. To see that, it is enough to integrate Eq. \eqref{eq:time-rel}, which leads to $T_A(t_B) = T_A(t_B^0) + (t_B - t_B^0) I$, where $t_B^0$ is some initial instant in clock $B$. Then, because the variance of the identity operator vanishes, we conclude that the variance of $T_A(t_B)$ equals the variance of $T_A(t_B^0)$ for every $t_B$.

For another way to reach the same conclusion, define $\Delta T_A^2 \equiv T_A^2 - \langle T_A\rangle^2 I$. Then, the Heisenberg equation implies that
\begin{equation}
    \Delta T_A^2(t_B) = \Delta T_A^2(t_B^0) + 2 (t_B - t_B^0) [T_A(t_B^0) - \langle T_A(t_B^0)\rangle I].
\end{equation}
Again, this allows us to conclude that the variance $\langle \Delta T_A^2\rangle$ does not change in time since the above equation leads to $\langle \Delta T_A^2(t_B)\rangle = \langle \Delta T_A^2(t_B^0)\rangle$. Even if we decide to consider the operator $\Delta T_A^2$ instead of its expected value, we can choose an initial state of clock $A$ with no initial uncertainty, so that we have $T_A(t_B^0) = \langle T_A(t_B^0)\rangle I$. In this case, not only the variance but also the operator $\Delta T_A^2$ does not change in $t_B$.

Observe that, instead of studying the modular energy in Eq. \eqref{mod-energy} that is typically discussed in the literature and refers to a variable associated with system $S$, we consider the modular energy of clock $A$. As already said, this clock plays the effective role of external systems interacting with $R$. Then, we \textit{infer} the nonlocality in time of the modular energy of $R$ from the nonlocality of the modular energy of $A$. More than that, we infer the modular energy of $S$. In fact, since $dH_{eff}^B/dt_B$ vanishes with our assumption that $H_{BS}(T_B)$ is null, we conclude that $e^{i H_{eff}^B \tau/\hbar}$ is conserved. Thus, a change in the modular energy of $A$ implies a change in the modular energy of $S$. Because of that, the parameter $\tau$ that appears in the definition of the modular energy of $A$ is taken with system $S$ in mind. While the parameter $\ell$ in the modular momentum was chosen to be the separation between the slits (the spatial periodicity) in the analysis of the double-slit experiment discussed in Section \ref{appendix}, the parameter $\tau$ in the modular energy of clock $A$ should be associated with a time periodicity of system $S$. This last conclusion follows, again, from the complete uncertainty relation, also discussed in that section.

Furthermore, the complete uncertainty relation also implies that if the system is observed with a projective measurement, its modular energy becomes completely uncertain. While in the case of the double-slit experiment the analogous result prevented information about a present potential to be acquired faster than light, here it blocks access to information about a future potential acting over the particle. Notably, both scenarios may be linked via a Lorentz boost.

It should also be noted the modular energy of clock $A$ commutes with the modular time given by $e^{2\pi i T_A/\tau}$, which is a variable that evolves locally in time since
\begin{equation}
    \frac{d}{dt_B} e^{2\pi i T_A/\tau} = -\frac{i}{\hbar} [e^{2\pi i T_A/\tau}, H_{eff}^B] = \frac{2\pi i}{\tau} e^{2\pi i T_A/\tau}.
\end{equation}
This, however, does not imply a violation of the uncertainty principle. A way to see that is by observing that the above result is proportional to the commutator $[e^{2\pi iT_A/\tau}, H_A]$ and, moreover,
\begin{equation}
    [e^{iH_A\tau/\hbar}, T_A] = \tau e^{iH_A\tau/\hbar}.
\end{equation}
This means that, while the modular variables of time and energy divides the time-energy phase space into disjoint cells and it is possible to simultaneously know the state of the system in each individual cell, the modular variables do not provide information about in which cell the state of the system is.

\section{Operational meaning of the modular energy in the timeless framework}
\label{sec:operational}

To give an operational meaning to the modular energy, we first obtain an analog of the expected value of the modular position given by Eq. \eqref{eq:expval-phase}. For that, we first observe that the expected value of any observable $O$ acting on $S$ at a given instant of time $t$ in clock $A$ can be computed in the timeless framework as the expected value of $|t\rangle \langle t| \otimes I_B \otimes O$ for the state $|\Psi\drangle$, i.e., $\langle O\rangle^t = \dlangle\Psi|\left(|t\rangle \langle t| \otimes I_B \otimes O \right)|\Psi\drangle$. Observe that we use the notation $\langle O\rangle^t$ instead of the standard $\langle O\rangle$ to evidence its dependence on $t$. Although this dependence is always present, it is not always acknowledged. This type of reasoning, without the inclusion of clock $B$, is also behind a recent definition of time of arrival \cite{maccone2020quantum}. Also, note that, in case $O$ is the modular momentum, Eq. \eqref{eq:expval-phase} is recovered.

The above idea is valid even if $O$ is also allowed to act in the space of clock $B$. However, if we consider the general case of $O$ acting on an arbitrary subsystem, which includes clock $A$, because of the possible lack of commutativity, there exist at least three possibilities for the definition of the expected value at a certain instant $t$ in clock $A$. In fact, omitting the tensor products and writing $\Pi_t=|t\rangle \langle t|$, such a definition can be, for instance, introduced with a symmetrization of the operators, as in
\begin{equation}
    \langle O\rangle_\text{sym}^t \equiv \frac{1}{2} \dlangle\Psi|\{\Pi_t, O\} |\Psi\drangle,
    \label{eq:symdef}
\end{equation}
where $\{\cdot,\cdot\}$ is the anticommutator. Moreover, one could introduce this definition without the symmetrization. In this case, we are left with either
\begin{equation}
    \langle O\rangle_\text{left}^t \equiv \dlangle\Psi|\Pi_t O |\Psi\drangle
\end{equation}
or
\begin{equation}
    \langle O\rangle_\text{right}^t \equiv \dlangle\Psi| O\Pi_t |\Psi\drangle.
    \label{eq:rigdef}
\end{equation}

Interestingly, there exists a discussion in the literature about the multiplicity of possibilities to define two-time correlators \cite{dolby2004conditional, giovannetti2015quantum, baumann2021generalized, trassinelli2022conditional}. Even though we want to define the expected value at a certain instant of time here, the lack of commutativity between $\Pi_t$ and $O$ implies that it will depend on the wave function at least for two points in time -- as given by clock $A$. As a result, these problems may bear some relation with each other.

The question we face here is, then, about how to choose which of the three definitions is the appropriate one. In principle, it seems that it is not possible to make such a choice. However, let us consider a scenario such that the state $\psi$ of system $R$ in Eq. \eqref{eq:history} is such that $|\psi(t_A + \tau)\rangle = e^{i\varphi} |\psi(t_A)\rangle$ for a certain time parameter $\tau$ and every $t_A$. Then, if we compute the expected value of the modular energy of system $R$, for which the three definitions are equivalent, we obtain $\langle e^{iH_R\tau}\rangle^t = \langle \psi(t+\tau)|\psi(t)\rangle$, i.e., it depends only on instants $t$ and $t+\tau$. Hence, the same should be expected for the modular energy of clock $A$ if we want to maintain the notion of conservation of this quantity. However, the definition in Eq. \eqref{eq:symdef} results in a dependence on $t-\tau$, $t$, and $t+\tau$, while the definition in Eq. \eqref{eq:rigdef} depends on $t-\tau$ and $t$. As a result, we are left with $\langle e^{iH_A\tau}\rangle^t \equiv \langle e^{iH_A\tau}\rangle_\text{left}^t$, which gives
\begin{equation}
    \begin{aligned}
        \langle e^{iH_A \tau/\hbar}\rangle^t &= \dlangle \Psi | \left(|t\rangle \langle t| e^{iH_A\tau/\hbar}\right) |\Psi\drangle \\
        &= \dlangle\Psi|\left(|t\rangle \langle t+\tau| \right) |\Psi\drangle \\
        &= \langle \psi(t)|\psi(t+\tau)\rangle \\
        &= e^{i\varphi}.
    \end{aligned}
    \label{eq:phase-a}
\end{equation}
This result is analogous to the one for the modular momentum in Eq. \eqref{eq:expval-phase}. There, however, the expected value had a half-factor multiplying the exponential. This is the case because the expected value computed there is associated with the relative phase between two halves of a (normalized) wave function at a single instant of time, while the computation done here captures a global phase difference between two wave functions, each at a different instant of time. Moreover, there exists a sign difference between the phases in Eq. \eqref{eq:phase-a} and $\langle e^{iH_R\tau}\rangle^t = e^{-i\varphi}$. This is associated with the conservation of the total modular energy --- in the Page and Wootters framework, the expected value for the energy (and the modular energy) of the entire system is zero.

The above discussion provides an operational meaning to the modular energy. Its expectation value can be obtained by performing two tomographies of system $S$: one at $t$ and the other at $t+\tau$. However, in some special cases, one may optimize this process. For instance, omitting clock $B$'s system, observe that for an eigenstate of energy $|E\rangle$, the expected value $\langle e^{iH_S \tau/\hbar}\rangle^t$ is just the dynamical phase $e^{iE_n\tau/\hbar}$. Then, if a system has a discrete energy spectrum and the state of $S$ is $|\psi\rangle = \sum_n c_n |E_n\rangle$, $\langle e^{iH_S \tau/\hbar}\rangle^t = \sum_n |c_n|^2 e^{iE_n\tau/\hbar}$. In this case, the tomography of $|\psi\rangle$, together with its spectral analysis, suffice for the measurement of $e^{iH_S\tau/\hbar}$. Also, in scenarios where the system can be placed in an interferometer, one may use a delay line in one arm, which results in a direct interference between $|\psi(t)\rangle$ and $|\psi(t+\tau)\rangle$.

Observe that the state of clock $B$ was omitted in this discussion and $|\psi\rangle$ was taken to be the state of system $S$. This can be done by assuming that clock $B$ starts disentangled to $S$. Then, the analog of Eq. \eqref{eq:time-rel} from the perspective of $A$, i.e., $dT_B/t_A = I$ implies that the two systems remain disentangled and, moreover, that the displacement of clock $B$'s state remains unchanged throughout the dynamics.

In what follows, we use the proposed formulation in order to analyze two scenarios where nonlocality in time seems to play a role.

\section{Applications}
\label{sec:applications}

\subsection{Particle and the piston}

\begin{figure}
    \centering
    \includegraphics[width=\columnwidth]{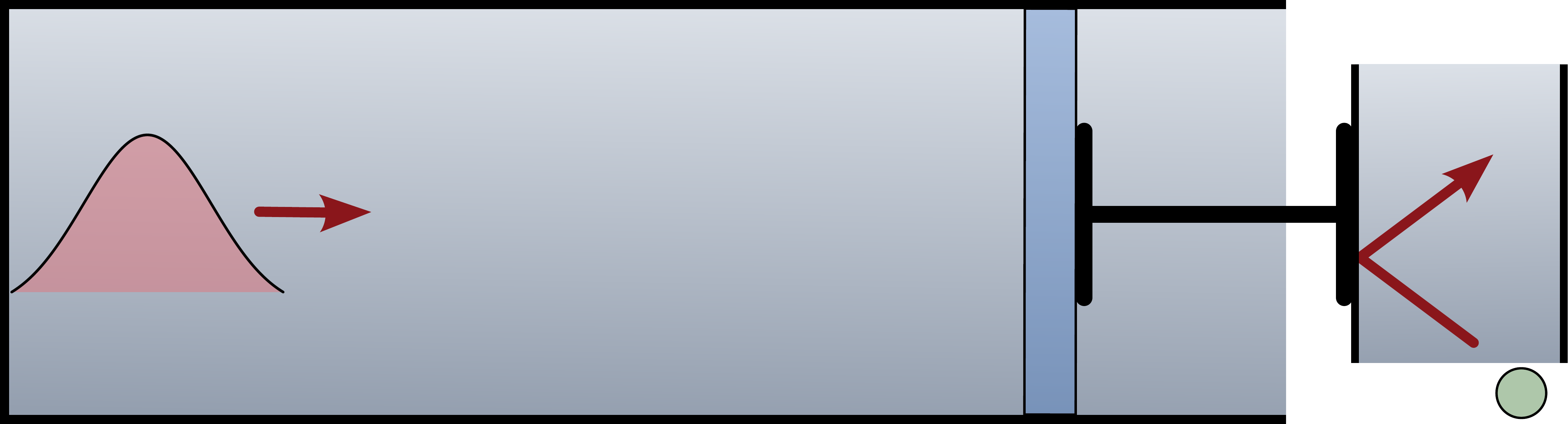}
    \caption{\textbf{Nonlocal interaction between a particle (pink wavepacket) in a box (bigger gray area) and a ball (green circle) outside of it.} The box's right-hand sidewall is a movable piston (blue region), which has a second box with an open top and bottom (smaller gray area) attached to it. The particle is in a periodic motion inside the closed box. While it is as far left as possible from the piston, the ball hits the open box twice, putting the piston in motion for a brief period of time. Although the particle and the box never get near each other, they exchange modular energy.}
    \label{fig:particle-piston}
\end{figure}

Now, we shall study a thought experiment previously suggested in Ref. \cite{aharonov2005quantum}. We start by introducing the standard description of it. For that, consider a long rectangular box with a movable piston on its right-hand side. Inside it, a quantum particle well-localized in a region much smaller than the length $\ell$ of the box is moving back-and-forth in periodic motion with period $\tau$ and negligible spreading. Also, assume a second box with two open sides is attached to the piston. This scenario is represented in Fig. \ref{fig:particle-piston}.

Then, suppose that the particle inside the closed box is located at its left end, when at time $\tau_1$, an external ball hits the box attached to the piston from its inside, putting the piston in a motion to the left. Later on, at an instant $\tau_2$, the ball hits the other inner side of the box, stopping the piston after it had moved a distance $\delta\ell$. It is assumed that $\tau_2-\tau_1\ll\tau$ in order to assure that the particle will remain distant from the piston during its translation.

Classically, it is expected that there will be no interaction between the particle and the ball. In a quantum treatment, however, it can be argued that there is an exchange (nonlocal in time) \footnote{We clarify that, although we refer to ``exchanges of modular energy,'' the conservation law for variables built from $e^{iH\tau/\hbar}$ is not additive. However, the reason the term ``exchange'' is used here (and in the literature at large) is that the conservation law for $H_{mod}$, defined as $H\mod 2\pi\hbar I/\tau$, is (modular) additive.} of modular energy between the two systems \cite{aharonov2005quantum}, and the explanation for that is the fact that, if $|\varphi(0)\rangle$ is the initial state of the particle, in cycles where the piston does not move
$|\varphi(\tau)\rangle = e^{i\alpha} |\varphi(0)\rangle$, where $\alpha\in[0,2\pi)$ and, hence, denoting the Hamiltonian of the particle by $H_S$, $\langle\varphi(0)| e^{iH_S\tau} |\varphi(0)\rangle = \langle\varphi(\tau) | \varphi(0)\rangle = e^{-i\alpha}$. However, if the ball hits the piston, the final position of the particle of a period $\tau$ is shifted by a distance $2\delta\ell$ with respect to its initial position. As a result, $|\varphi(\tau)\rangle = e^{i(\alpha+2P\delta\ell/\hbar)} |\varphi(0)\rangle$ and $\langle\varphi(0)| e^{iH_S\tau} |\varphi(0)\rangle \approx e^{-i(\alpha+2\langle P\rangle \delta\ell/\hbar)}$. Hereon, we show how to analyze this example within the timeless framework.

The time-independent Hamiltonian of the particle inside the closed box, i.e., its Hamiltonian in case the external ball does not interact with the piston can be written as
\begin{equation}
    H_S = \frac{1}{2m} P^2 + V_l(X) + V_r(X),
\end{equation}
where $V_l$ and $V_r$ are the potentials associated respectively with the left and the right walls of the box. For simplicity, one could take $V_r(x)\equiv V_l(x-\ell)$.

Now, since instants $\tau_1$ and $\tau_2$ refer to events related to external systems, we assume they are observed in clock $A$. Then,
\begin{equation}
    H_{int}(T_A) = V_r(X+f(T_A)) - V_r(X),
\end{equation}
where
\begin{equation}
    f(t) = \delta\ell \left[ \frac{t-\tau_1}{\tau_2-\tau_1} \Theta(t-\tau_1)+ \frac{\tau_2-t}{\tau_2-\tau_1}\Theta(t-\tau_2)\right]
\end{equation}
and $\Theta$ is the Heaviside step function.

With that, the Hamiltonian $H_{eff}^B$ in Eq. \eqref{ta-hamiltonian} becomes
\begin{equation}
    H_{eff}^B = H_A + \frac{1}{2m} P^2 + V_l(X) + V_r(X+f(T_A))
\end{equation}
and, hence, it follows from Eq. \eqref{mod-energy-nonlocality} that, over the cycle and, in special, the first half of the cycle in which the piston had its position changed,
\begin{equation}
    \frac{d}{dt_B}e^{iH_A\tau/\hbar} = -\frac{i}{\hbar} \left[V_r(X+\delta\ell I) - V_r(X)\right] e^{iH_A\tau/\hbar}.
    \label{eq:result-ex1}
\end{equation}

Observe that, in our treatment, instead of including the dynamics of the external ball, its effect (i.e., the change in position of the piston) was included as a time-dependent potential. This is similar to other approaches usually considered in the literature where external effects are represented by potentials in the Hamiltonian. Then, Eq. \eqref{eq:result-ex1} implies that, even while the particle is as far away inside the box from the piston as possible, there exists an exchange of modular energy between the exterior (represented by clock $A$) and system $R$ that depends on the final position of the piston. However, this exchange does not lead to superluminal communication because, as the complete uncertainty relation implies, an observation of the particle or the ball would make their modular energy completely uncertain.

\subsection{Nonlocal interaction in time between magnetic fields and a spin}

The scenario that will be considered in this section was the first example of a nonlocal exchange of modular energy discussed in the literature. It was introduced in the seminal article \cite{aharonov1969modular} and later revisited in Ref. \cite{aharonov2005quantum}. Our aim here is to characterize the nonlocality in time associated with this exchange in the timeless framework.

\begin{figure}
    \centering
    \includegraphics[width=\columnwidth]{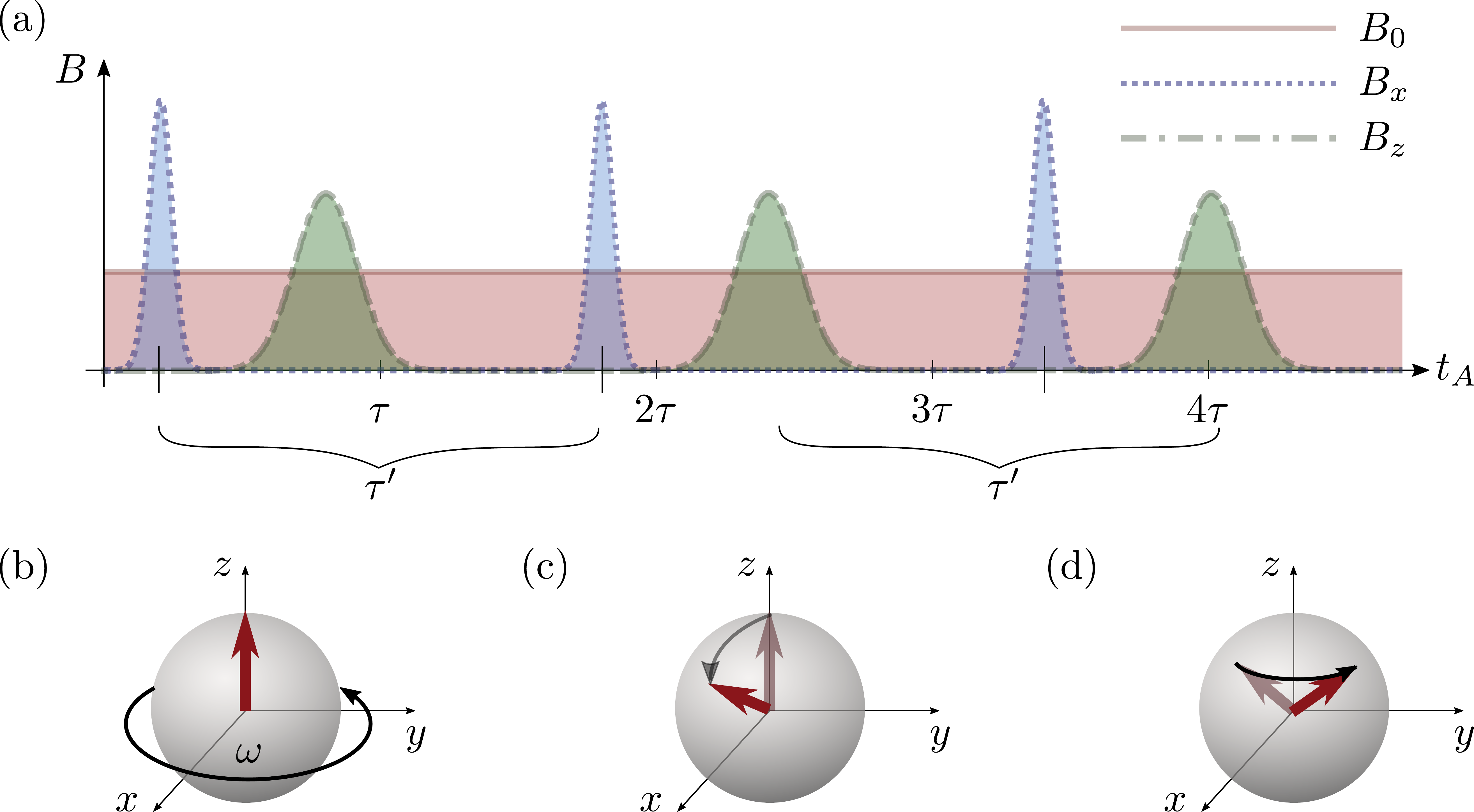}
    \caption{\textbf{Representation of a thought experiment with magnetic fields interacting nonlocally in time and affecting the dynamics of a spin, initially prepared in the state $|1\rangle$.} (a) The magnetic fields applied to the system are displayed as a function of time, as seen from an external clock. (b) First, a constant magnetic field $B_0$ is applied in the $z$ direction, generating a rotation of the spin about its $z$ axis with period $\tau$. (c) Then, periodic pulses of magnetic field $B_x$ in the $x$ direction with period $\tau'\neq n\tau$, where $n\in\mathbb{Z}$, are also applied upon the spin, generating partial rotations around its $x$ axis. Since $\tau'$ is not a multiple of $\tau$, the spin does not flip. (d) However, the introduction of extra pulses of magnetic field $B_z$ in the $z$ direction with period $\tau'$ can be designed in order to flip the spin, even if the pulses in the $x$ and $z$ directions never coincide in time.}
    \label{fig:magnetic-fields}
\end{figure}

To start, consider a spin-$1/2$ particle with magnetic moment $\mu$ under the influence of a constant magnetic field $B_0$ in the $z$ direction. Then, its Hamiltonian is
\begin{equation}
    H_S = \frac{\hbar}{2}\mu B_0 \sigma_z.
\end{equation}
In this case, the spin dynamics is simply given by a rotation around the $z$ axis with angular frequency $\omega = \mu B_0$. This scenario is presented in Fig. \ref{fig:magnetic-fields}(a).

Now, suppose that an additional magnetic field $B_x(t)$ consisting of periodic pulses with period $\tau'$ is added to the dynamics of the spin, as shown in Fig. \ref{fig:magnetic-fields}(b). In this case, it can be asked whether a spin that starts in an eigenstate of $\sigma_z$ direction flips or not. To answer this question, one needs to observe that, under $H_S$, the system completes each cycle in a period $\tau=2\pi/\omega = 2\pi/\mu B_0$. Then, with the addition of $B_x(t)$, if $\tau'$ is a multiple of that, i.e., $\tau'=n\tau$ for a non-zero integer $n$, the spin will eventually flip. Otherwise, the effect of $B_x(t)$ vanishes on average and the spin remains unchanged.

Assume, then, that $\tau'\neq n\tau$ and apply an additional periodic pulse of magnetic field in the $z$ direction $B_z(t)$, also with a period $\tau'$, as illustrated in Fig. \ref{fig:magnetic-fields}(c). The pulses $B_x(t)$ and $B_z(t)$ are such that their product is $B_x(t)B_z(t)=0$ for every $t$, i.e., they are not applied at the same instant of time. Now, it becomes possible to flip the spin again. In fact, $B_z(t)$ can be conveniently chosen in such a way that the effective rotation about the $z$ axis caused by $B_0$ and $B_z(t)$ has a period $T$ such that $\tau=mT$ for a non-zero integer $m$.

As can be understood from this description, and pointed out in Ref. \cite{aharonov2005quantum}, the analysis of the precession of the spin has a classical analog. However, the change of energy associated with the flip of the spin seems to be puzzling from a classical perspective. In fact, in Ref. \cite{aharonov2005quantum}, after careful consideration of the problem, the authors concluded that, while making no net contribution to the energy, the field $B_z(t)$ seems to modify the way the particle and $B_x(t)$ exchange energy, allowing exchanges that are not multiples of $2\pi\hbar/\tau'$. Because of that, it is commonly claimed that the exchange of energy that causes the flip of the spin is nonlocal in time.

Here, the aim is to formalize this idea in the timeless framework. To start, observe that the time-dependent part of the evolution is described by the Hamiltonian
\begin{equation}
    H_{int}(T_A) = \frac{\hbar}{2}\mu \left[B_x(T_A) \sigma_x + B_z(T_A)\sigma_z\right].
\end{equation}
Then, the effective Hamiltonian $H_{eff}^B$ is
\begin{equation}
    \begin{aligned}
    H_{eff}^B &= H_A + H_S + H_{int}(T_A) \\
    &= H_A + \frac{\hbar}{2}\mu \left[B_x(T_A) \sigma_x + (B_0 + B_z(T_A))\sigma_z\right]
    \end{aligned}
\end{equation}
and
\begin{equation}
    \frac{d}{dt_B}e^{iH_A\tau'/\hbar} = 0.
\end{equation}
However, $\tau'$ is a frequency associated with the external system, and it is being assumed that there exists no integer $n$ such that $\tau'=n\tau$, i.e., $\tau'$ is not a multiple of the ``natural'' period $\tau$ associated with the spin. Hence, the dynamics of the modular energy of interest is
\begin{equation}
    \begin{aligned}
    \frac{d}{dt_B}e^{iH_A\tau/\hbar} = -\frac{i\mu}{2} \left[\right. &B_x(T_A+\tau) \sigma_x + B_z(T_A+\tau I) \sigma_z - \\
    &\left. - B_x(T_A) \sigma_x - B_z(T_A) \sigma_z \right] e^{iH_A\tau/\hbar}.
    \end{aligned}
    \label{temp-eq-dyn}
\end{equation}
Assuming that each pulse associated with $B_x(t)$ and $B_z(t)$ has a sufficiently short duration, Eq. \eqref{temp-eq-dyn} leads to
\begin{equation}
    \frac{d}{dt_B}e^{iH_AT/\hbar} = -\frac{i\mu}{2} \left[B_x(T_A+\tau I) \sigma_x - B_z(T_A) \sigma_z \right] e^{iH_A\tau/\hbar}
\end{equation}
if $B_x(T_A)=0$, and
\begin{equation}
    \frac{d}{dt_B}e^{iH_A\tau/\hbar} = -\frac{i\mu}{2} \left[B_z(T_A+\tau I) \sigma_z - B_x(T_A) \sigma_x\right] e^{iH_A\tau/\hbar}
\end{equation}
if $B_x(T_A+\tau I)=0$.

Those equations show how the modular energy of the system has its evolution affected simultaneously by the presence of the fields $B_z$ and $B_x$ -- even though they do not occur simultaneously in time. This nonlocal interaction in time between the fields and the spin is what modifies how they exchange energy.

\section{Discussion}
\label{sec:discussion}

We have developed and applied a formalization of modular energy within the timeless framework of quantum mechanics. This puts modular energy and time on an equal footing with the modular position and momentum. While our approach helps to clarify how dynamical nonlocality in time is present in the study of modular energy, our results might be just the starting point for a complete comprehension thereof.

For instance, like the modular position and momentum divide the phase space into periodic cells, modular energy and its associated modular time also divide the energy-time phase space in this way. From the conceptual point of view, the meaning of such a division may deserve further clarification. This could have ramifications in other areas of quantum mechanics, like time crystals, where the evaluation of the modular energy corresponds to the Floquet quasi-energy.

Moreover, if an ultimate Planck scale limit on time is assumed, the framework developed here can be particularly helpful since the modular energy promotes translations of length $\tau$ in time. Therefore, with the appropriate parameter $\tau$, this modular variable and the ensuing nonlocality in time could play a role in our understanding of spacetime. More precisely, modular energy as formalized here may be an important tool in understanding recent studies in the direction of modular spacetime \cite{freidel2016quantum, yargic2020path}.

Finally, quantum events with indefinite causal order have been attracting some attention lately \cite{oreshkov2012quantum, brukner2014quantum, rubino2017experimental, goswami2018indefinite, barrett2019quantum, barrett2021cyclic}. Although the connections between these studies and the present work still need to be examined, causal structures implicitly involve both the concepts of kinematics and dynamics, possibly even replacing them in future approaches to physics \cite{spekkens2015paradigm}. Nevertheless, meanwhile, it seems that modular energy, and its nonlocality in time, may bring new insights into such scenarios.

\acknowledgements{We are grateful to Yakir Aharonov for many helpful discussions. In addition, we wish to thank Leon Bello, Jordan Cotler, Pawe\l{} Horodecki, Zohar Schwartzman-Nowik, and Mordecai Waegell for their constructive comments regarding the current text. This research was supported by grant number FQXi-RFP-CPW-2006 from the Foundational Questions Institute and Fetzer Franklin Fund, a donor-advised fund of Silicon Valley Community Foundation, by the Israeli Innovation authority (Grants 70002 and 73795), by the Pazy foundation, and by the Quantum Science and Technology Program of the Israeli Council of Higher Education.}

\bibliography{citations}

\end{document}